\documentclass{WileyMSP-template}
\usepackage{amsmath}
\usepackage[mathlines]{lineno}
\usepackage{amssymb}
\usepackage{url}
\usepackage{xcolor}
\usepackage[utf8]{inputenc}
\begin{document}

\title{Position-controlled functionalization of vacancies in silicon by single-ion implanted germanium atoms}

\maketitle


\author{Simona Achilli*}
\author{Nguyen H. Le}
\author{Guido Fratesi}
\author{Nicola Manini}
\author{Giovanni Onida}
\author{Marco Turchetti}
\author{Giorgio Ferrari}
\author{Takahiro Shinada}
\author{Takashi Tanii}
\author{Enrico Prati}



\begin{affiliations}
S. Achilli, G. Fratesi, N. Manini, G. Onida\\
Dipartimento di Fisica, Universit\`a degli Studi di Milano, Via Celoria 16, 20133 Milano, Italy.\\
Email Address: simona.achilli@unimi.it

Nguyen H. Le\\
Advanced Technology Institute and Department of Physics, University of Surrey, Guildford GU2 7XH, United Kingdom.\\

Marco Turchetti\\
Istituto di Fotonica e Nanotecnologie, Consiglio Nazionale delle Ricerche, Piazza Leonardo da Vinci 32, 20133 Milano, Italy.\\
Current affiliation: Research Laboratory of Electronics, Massachusetts Institute of Technology, Cambridge, MA 02139, USA.\\

Giorgio Ferrari\\
Dipartimento di Elettronica, Informazione e Bioingegneria, Politecnico di Milano, Piazza Leonardo da Vinci 32, 20133 Milan, Italy

Takahiro Shinada\\
Center for Innovative Integrated Electronic Systems (CIES), Tohoku University, 468-1 Aramaki Aza Aoba, Aoba-ku, Sendai, Miyagi, 980-8572, Japan\\

Takashi Tanii\\
Faculty of Science and Engineering, Waseda University, 3-4-1 Ohkubo, Shinjuku, 169-8555 Tokyo, Japan.\\

Enrico Prati\\
Istituto di Fotonica e Nanotecnologie, Consiglio Nazionale delle Ricerche, Piazza Leonardo da Vinci 32, 20133 Milano, Italy.\\
Email Address: enrico.prati@cnr.it

\end{affiliations}





\begin{abstract}

Special point defects in semiconductors have been envisioned as suitable components for quantum-information technology.

The identification of new deep centers in silicon that can be easily activated and controlled is a main target of the research in the field.
Vacancy-related complexes are suitable to provide deep electronic levels but they are hard to control spatially.

With the spirit of investigating solid state devices with intentional vacancy-related defects at controlled position, here we report on the functionalization of silicon vacancies by implanting Ge atoms
through
single-ion implantation,
producing Ge-vacancy (Ge$V$) complexes.
We investigate the quantum transport through an array of Ge$V$ complexes in a silicon-based transistor.

By exploiting a model based on an extended Hubbard Hamiltonian derived from {\it ab-initio} results 
we find anomalous activation energy values of the thermally activated conductance of both quasi-localized and delocalized many-body states, compared to conventional dopants. 
We identify such states, forming the upper Hubbard band, as responsible of the experimental sub-threshold transport across the transistor.

The combination of our model with the single-ion implantation method enables future research for the engineering of Ge$V$ complexes towards the creation of spatially controllable individual 
defects in silicon for applications in quantum information technologies. 
\end{abstract}



\section{\label{sec:level1}Introduction}

Deep-level impurities in group IV semiconductors have been envisioned as a possible route toward room-temperature quantum information processing \cite{Ono19,belli2014electron,achilli-sci-rep,bockstedte2018ab}. 
 Vacancies in silicon could in principle
 provide deep electrically relevant individual centers suitable
 for high temperature quantum information manipulation
 \cite{shinada2016deterministic,shinada2014opportunity}, but their positioning is rarely controllable \cite{schofield2013quantum}.
Indeed, the miniaturization of  nanoelectronic devices for quantum information processing \cite{rotta2017quantum} faces many technological and fundamental challenges, such as the control of effects induced by structural disorder \cite{Wacquez2010} and those related to the localization
of the dopant wavefunction in nanometer-size channels \cite{moraru2011atom}.
While the former can be partly addressed by exploiting high-precision implantation techniques \cite{van2015single,hori2012quantum} or scanning-tunneling microscopy \cite{oberbeck2004measurement}, the wavefunction localization
could
be a major issue, due to the crucial role it 
plays
in phenomena related to electronic correlation, transport, and eventually in setting up
controlled
electronic properties
suitable for exploitation
in quantum information devices.

Quantum transport in silicon field-effect transistor with As, B, P dopants has been
probed both at cryogenic and room temperature \cite{prati2016band, fuechsle2012single, Wang16,hamid2013electron}.
By controlling the dopant density with atomic precision
it is possible to explore different quantum conduction regimes \cite{leti2011switching, shinada2011quantum} ranging from single-electron Coulomb blockade to Hubbard-band transport
in the Anderson-Mott transition
region
\cite{prati2012anderson,prati2016band}.
Nevertheless the application of conventional dopants as qubits requires operating temperature below 1.5~K \cite{petit2020universal},
preventing the exploitation of this technology between soft cryogenic and room temperatures.

Such drawback can be mitigated by exploiting vacancy-related complexes that, by virtue of their deep electronic levels, could be used for quantum transport at room temperature.
The functionalization through foreign species would also allow a spatial control of these defects and could be exploited to induce a spatial anisotropy of the electronic wavefunctions without altering the deep electronic spectrum, leading to a reduction of the interference effects \cite{Wang16}.
Fast optical manipulation like NV centers and divacancies in SiC enable in principle up to room temperature operations \cite{bockstedte2018ab}.
Moreover, tailoring correlation effects in nanostructure via a precise control of both the positioning and the composition of point defects would open the way to {\it quantum metamaterials.} 
This challenge benefits of the interplay between advanced experimental techniques and theoretical models able to predict the
properties
of interest
\cite{Narang19}.

Based on {\it ab-initio} calculations \cite{achilli-sci-rep}, 
some of us have
recently demonstrated that Ge$V$ complexes in silicon are stable defects, characterized by excited states deep in the bandgap, at about $\simeq -0.5$~eV and $\simeq -0.35$~eV from the conduction band, consistent with
experiment \cite{Supr95}.
Such impurity states have large on-site electron-electron repulsion ($\simeq150$~meV)  
due to their highly localized wavefunctions (the decay length of the
lowest
charged
state is $\simeq 0.45$~nm).
This localization far exceeds that of conventional P and B states \cite{Smit17}. 
Here we exploit this theoretical prediction
for the fabrication of
the first transistor with Ge$V$ complexes in the silicon channel,
and we interpret the electric conductance measurements
by means of a Hubbard-like model.

In view of addressing the challenge of functionalizing vacancies in silicon as deep-level spatially-controlled centers, we generate Ge$V$ complexes by single-ion implantation (SII) of germanium atoms.
The overall strategy consists in investigating the trans-conductance of silicon transistors of length of both 200~nm and 500~nm, operated from cryogenic to room temperature. The transistor's channel consists of a linear array of up to 100 implanted Ge atoms, which have been activated at the moderate annealing temperature of $550^\circ$C.

The quantum transport across an array of such defects reveals the formation of impurity bands inside the silicon band gap.
Differently from similar experiments carried out with
phosphorous impurities, in which activation energies for thermally activated quantum transport are of the order of few meV \cite{prati2012anderson},
Ge$V$ complexes show different kinds of thermally-activated processes of the Hubbard band, that are characterized by activation energies ranging from hundreds of $\mu$eV to tens of meV.

For the modeling of a one-dimensional array of few Ge$V$ centers with linear
density and disorder comparable to experiments,
we exploit an extended Hubbard approach.
The model exploits Hamiltonian parameters derived from the {\it ab-initio} wavefunctions instead of a Kohn-Luttinger description \cite{Shindo,Saraiva,Dusko}, going beyond an effective-mass theory approach \cite{Wella05}.
Through this formalism we shed light on the mechanisms of quantum transport in such
array of Ge$V$ complexes, in comparison to conventional dopants.

A side-by-side analysis
of the experimental and model data 
allows us to explain the observed impurity band as due to the upper Hubbard band.
By raising temperature from $4.2$~K up to room temperature, we demonstrate that the conductance is
driven by {\em weakly localized} states
at low temperature, while at high temperature
{\em delocalized} states start to contribute.

The spatially deterministic implantation of up to 100 Ge$V$ centers and electrical readout in our experiment 
may be further developed in the future to achieve, - thanks to lower implantation energies \cite{van2015single}, the sufficient spatial control to address for instance
large scale analog quantum simulation of strongly-correlated models with deep impurities in silicon. 
This system size is far beyond the capability of
conventional-computer
exact simulations and is thus the regime of interests for future semiconductor-based quantum simulators \cite{salfi2016quantum,hensgens2017quantum}.
Although it is not granted that a nanowire of 100 implanted Ge$V$ sites may eventually realize a quantum simulator with 100 qubits,
impurities in silicon complement cold-atom platforms \cite{Greif2013,Esslinger2010}, as they enable us to access interaction strengths and hopping amplitudes that are
orders of magnitude higher, easing the need for
extreme
cooling \cite{salfi2016quantum}.
Ge$V$ is also deeper than shallow dopants, with the binding energy of the double occupancy state more than 200 times the $1.7$~meV value in Si:P \cite{Saraiva}, allowing in principle a much better protection against ionization to the conduction band.
Our transport measurements allow us to confirm starting fabrication parameters which could be further developed in the future to realize a quantum device based on Ge$V$ impurities.

\section{Results}\label{results:sec}

\subsection{Optimization of the linear implantation density}

As our approach
relies on measuring the conductance provided by the electrons hopping from one localized state to the next,
the device must attain a large enough linear density of Ge$V$ sites for ``impurity'' bands to form.
Thus, compared to the wide impurity states of As and P \cite{prati2016band},
the implantation pitch for Ge must be adapted to the size of the Ge$V$ localized states.
To assess  this size, we resort to {\it ab-initio}
calculations.

The Ge$V$ complex is a neutral defect, being Ge isoelectronic to silicon, and, differently from conventional dopants, does not show hydrogen-like shallow states, but deep energy levels related to the dangling bonds
in the vacancy.
In the Ge$V$ ground state a deep level within the valence band and a defect state in the gap are filled by the four unpaired electrons from the dangling bonds.
Further electrons,
injected into the channel
by  applying a small bias voltage between the source and the drain electrodes,
occupy strongly-bound
defect states in the energy gap, experiencing a relatively large on-site Coulomb repulsion ($U\simeq 150$ meV).
Depending on the amplitude of the inter-site hopping,
the channel may behave as a
more-or-less
strongly-correlated system, where the Mott-Hubbard physics applies \cite{Mott,Essler05}.

We model the one-dimensional Ge$V$ array with an extended Hubbard Hamiltonian \cite{Le,prati2012anderson,vanDongen}
\begin{align} \label{model:eq}
H=&\sum_i\epsilon_i n_i -\sum_{\langle i,j\rangle} \sum_{\sigma=\uparrow, \downarrow} t_{i,j}(c^\dagger_{i,\sigma}c_{j,\sigma}+c^\dagger_{j,\sigma}c_{i,\sigma})
\\
&+ U \sum_i n_{i,\uparrow}n_{i,\downarrow}+ \sum_{i\ne j} W_{i,j} n_i n_j
\,.
\end{align}
which includes the on-site energy level $\epsilon_i$ for each defect, the coupling $t_{i,j}$ between nearest-neighboring sites $\langle i,j\rangle$, the on-site Hubbard correlation $U$ energy, plus inter-site long-range electron-electron repulsion $W_{i,j}$. The $c_{i,\sigma}$ is a second-quantization fermionic operator which destroys a spin-$\sigma$ electron at site $i$;
$n_{i\sigma} = c^\dagger_{i,\sigma}c_{i,\sigma}$, and $n_i =  \sum_\sigma n_{i\sigma}$ are the number operators.
The screening by the host silicon matrix is accounted for, scaling $W_{i,j}$ by the dielectric constant of silicon ($\epsilon_{\rm Si}=11.7)$.
For further details, see Section~\ref{metods:sec}.

\begin{figure}
\centering
\includegraphics[width=\textwidth]{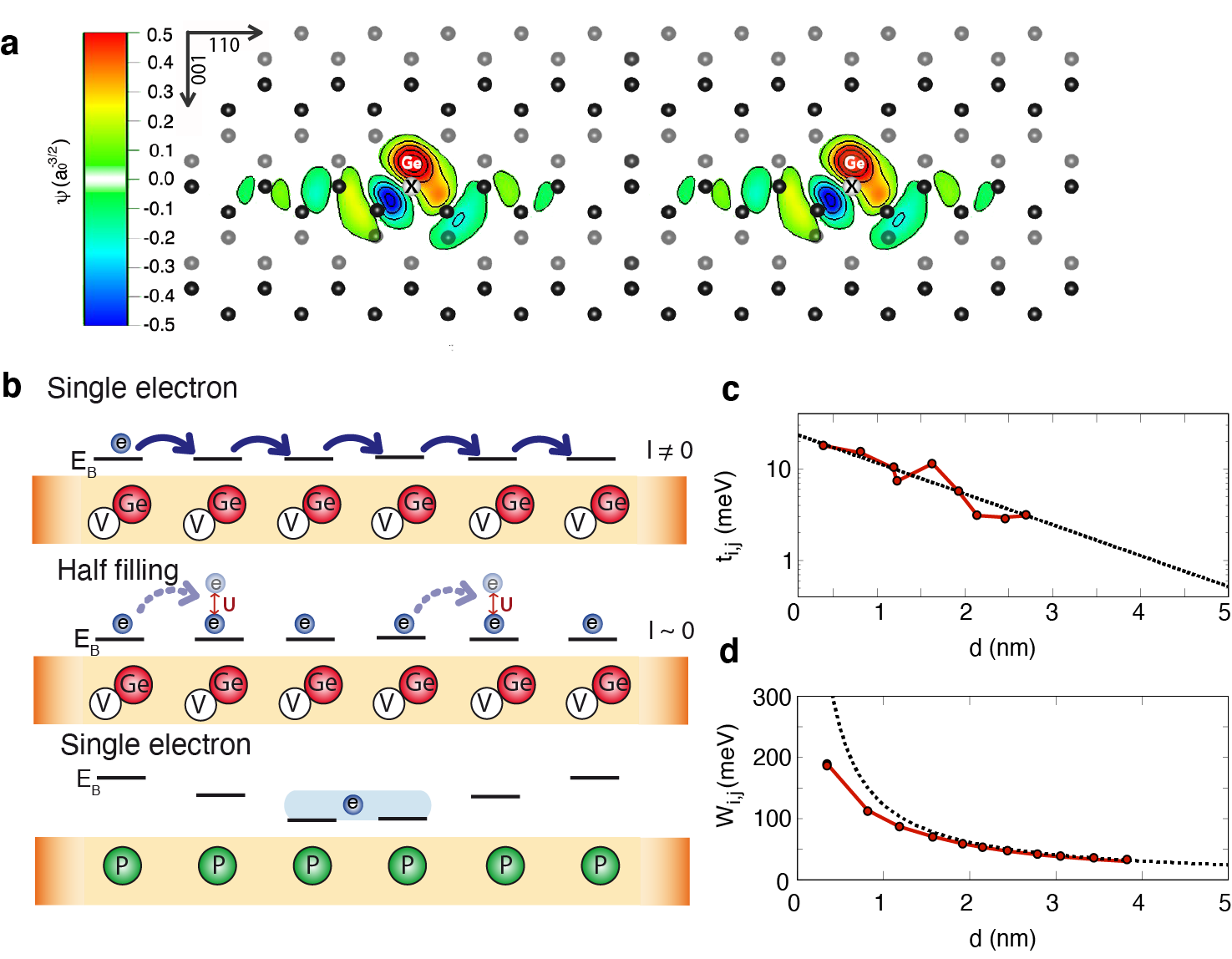}
\caption{\label{wavef}
(a) Wavefunction of the lowest charged state bound to the Ge$V$ complex, $-0.5$~eV below the conduction band.
Two complexes are aligned in the $[110]$ direction through the vacancy center (black crosses), and the panel reports the cut through a $[\overline{1}01]$ plane.
Si ions in the outermost crystal layer are indicated by black dots; 
gray dots mark ions in the second layer.
The 
Ge atom is out of plane. 
(b) The conduction mechanism through the array of Ge$V$ defects in the low-filling (top) and half-filling (middle) regime. The comparison with the case of P impurities in silicon is reported (bottom) showing the localization of the single electron wavefunction due to the different on-site energies of the impurities.
For Ge$V$, the single-electron and half-filling configurations are reported illustrating the effect of the on-site Hubbard correlation.
(c) Computed hopping parameter $t_{i,j}$ (dots) as a function of the inter-site distance.
The dotted line is the fitted exponential behavior.
(d) The electron-electron interaction obtained from the {\it ab-initio} wavefunction (dots).
The dotted line indicates the large-distance asymptotic decay
(see text).
}
\end{figure}

For charged defects such as donor or acceptor impurities, the on-site energy $\epsilon_i$ must include the long-range Coulomb interaction $V$ from the ion core of nearby sites \cite{Le}, giving rise to position-dependent on-site energies
$\epsilon_i$ (see bottom of Figure~\ref{wavef}b).
Since Ge$V$ is a neutral object, here the on-site energies are the same for all the sites in the array, ($\epsilon_i = \epsilon$, see Figure~\ref{wavef}b).
Such difference between Ge$V$ and conventional dopants leads to a different shape of the conductance spectrum, as discussed below.

We evaluate the hopping and inter-site repulsion parameters using the wavefunction of the electronic state occupied by an additional electron on the defect, namely the charged excitation sitting $-0.5$~eV below the conduction-band bottom \cite{achilli-sci-rep}.
The use of a wavefunction obtained from {\it ab-initio} calculation allows a more accurate description of deep energy levels of Ge$V$ with respect to the usually adopted Kohn-Luttinger wavefunction which is suitable for shallow levels of conventional dopants but not appropriate for Ge$V$.
Figure~\ref{wavef}a shows
a section of such  {\it ab-initio}-computed wave function across a high-symmetry crystal plane through two vacancy centers.
In the model, the Ge atoms bound to the vacancies lie along the [111] crystal direction (outside the plane reported in Figure~\ref{wavef}a).
The wavefunction in the proximity of the defect is asymmetric, with a larger amplitude at the Ge side of the vacancy, and exhibits
a fast decay along the [110] crystal direction that we chose as direction for implantation.
In real samples one can expect the Ge ions to randomly occupy one of the four substitutional positions around the vacancy. 
Nevertheless, due to their fast decay, defect wavefunctions at the  typical implantation distances are essentially isotropic.  We can hence expect that the overlap between wavefunctions at different sites is weakly affected by the  orientations of the Ge$V$ complexes and by the
Ge$V$-Ge$V$ joining line.

Figures~\ref{wavef}c and  ~\ref{wavef}d report the hopping and inter-site repulsion parameters computed with the {\it ab-initio} wavefunction of Figure~\ref{wavef}a.
The hopping $t_{ij}$ follows
approximately
an exponential decay as a function of distance,
with smaller
oscillations than typically observed for P in silicon.
Such oscillations are usually more marked at small distances, while they are quenched for large inter-site separations.
Accordingly,
for $t_{ij}$ we take
the exponential function of distance (black line in Figure~\ref{wavef}c), fitted on the {\it ab-initio} overlaps evaluated for $d<3$~nm. 

At all distances the hopping parameter is significantly smaller
than those relevant for conventional dopants \cite{Le} such as P, due to the stronger localization of the Ge$V$ wavefunction.
Guided by Figure~\ref{wavef}c, the Ge implantation pitch needs to be of order $d\simeq 5$~nm, for $t_{i,j}$ to reach an amplitude that supports measurable conductance in conventional P-doped single-impurity transistors, namely in the region around $t_{i,j}\simeq 1$~meV.
In our setup, we have implanted the ions at a $10$~nm spacing, compared to the $\sim 100$~nm spacing used for P \cite{prati2016band}.

The inter-site electron-electron repulsion $W_{ij}$ decays as
$V_0/d_{ij}$ (with $V_0=e^2/4 \pi \epsilon_0\epsilon_{\rm Si}$) 
for $d_{ij}\gtrsim 2$~nm (black curve in Figure~\ref{wavef}d).
Given the adopted spacing in the SII fabrication process,
for our calculations we stick to the $W_{ij}= V_0/d_{ij}$ asymptotic formula, namely the same expression used for other kinds of dopants.

\subsection{Thermally-activated transport in a disordered 1D array of Ge$V$ defects}

\begin{figure}
\includegraphics[width=0.9\textwidth]{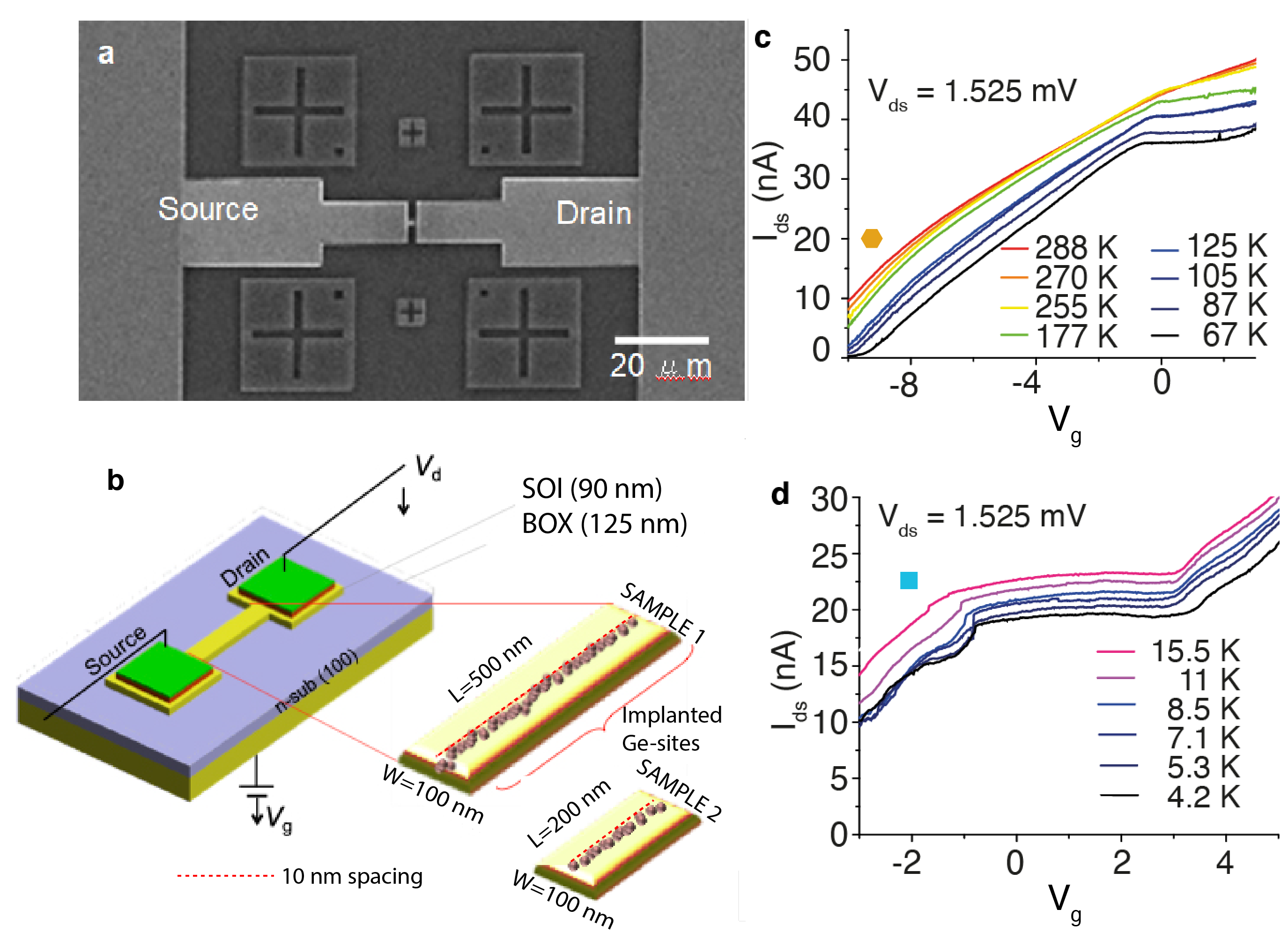}
\caption{\label{Fig-exp1}
(a) A scanning microscopy image and 
(b)~a scheme of the Ge-implanted transistor.
Two different lengths of the silicon channel are considered: 500 nm or 200 nm with 50 or 8 pairs of Ge ions, respectively, implanted by single-ion implantation.
The silicon-on-insulator (SOI) thickness is 90 nm
and the underlying BOX is of 125 nm.
(c) $I$--$V$ trans-characteristics, measured from  67 K to room temperature.
The array shows impurity band transport up to room temperature. The yellow hexagon marks the bias voltage corresponding to low-filling regime at which the activation energies are extracted  (see Figure \ref{disord}c).
(d) Low-temperature $I$--$V$ trans-characteristics.
A thermally activated band is formed below the conduction-band edge at +3.5~V.
The cyan square marks the bias voltage corresponding to low-filling regime at which the activation energies are extracted (see Figure \ref{disord}d). 
}
\end{figure}

We fabricated a set of nanotransistors doped by single-ion implantation \cite{shinada2014opportunity} whose channel charge
is controlled by a backgate, so the implantation is carried out
from the top.  
The channel of the transistor is either $200$~nm or $500$~nm long, depending on the sample. Eight and fifty pairs of Ge ions are implanted in the former and the latter, respectively (Figure~\ref{Fig-exp1}b) to form a 1-dimensional (1D) array connecting the source and the drain electrodes, as detailed in
Section~\ref{exp:sec} below.

The $I$--$V$ trans-characteristics (reported in Figure~\ref{Fig-exp1}c-d and Figure~\ref{short}b for the long and short channel transistor, respectively) shows that below the threshold of the transistor additional transport is observed after the implantation at sufficiently high density, due to impurity-band formation.
We have explored temperatures ranging between 4.2~K and 90~K for the short-channel sample (see also Supporting Information) while for the long one we extend up to room temperature.

A thermally activated impurity band is observed up to room temperature, whose peculiar activation energy values are discussed below.

In order to characterize the impurity-band transport, we simulate
the conductance by exploiting the rate-equation formalism \cite{Beenakker,Chen,Le}.
This method focuses on the tunneling of electrons from the leads into the array, leading to a change of the many-body state from $(n_e-1,\alpha)$ to $(n_e,\beta)$, where $n_e$ is the number of electrons in the array and $\alpha,\beta$ label the states.
A transition rate equation describes the tunneling of one electron from the left electrode into the array and from the array into the right electrode and a temperature dependent linear response conductance is obtained, see Section~\ref{metods:sec}.

\begin{figure}
\includegraphics[width=1.0\textwidth]{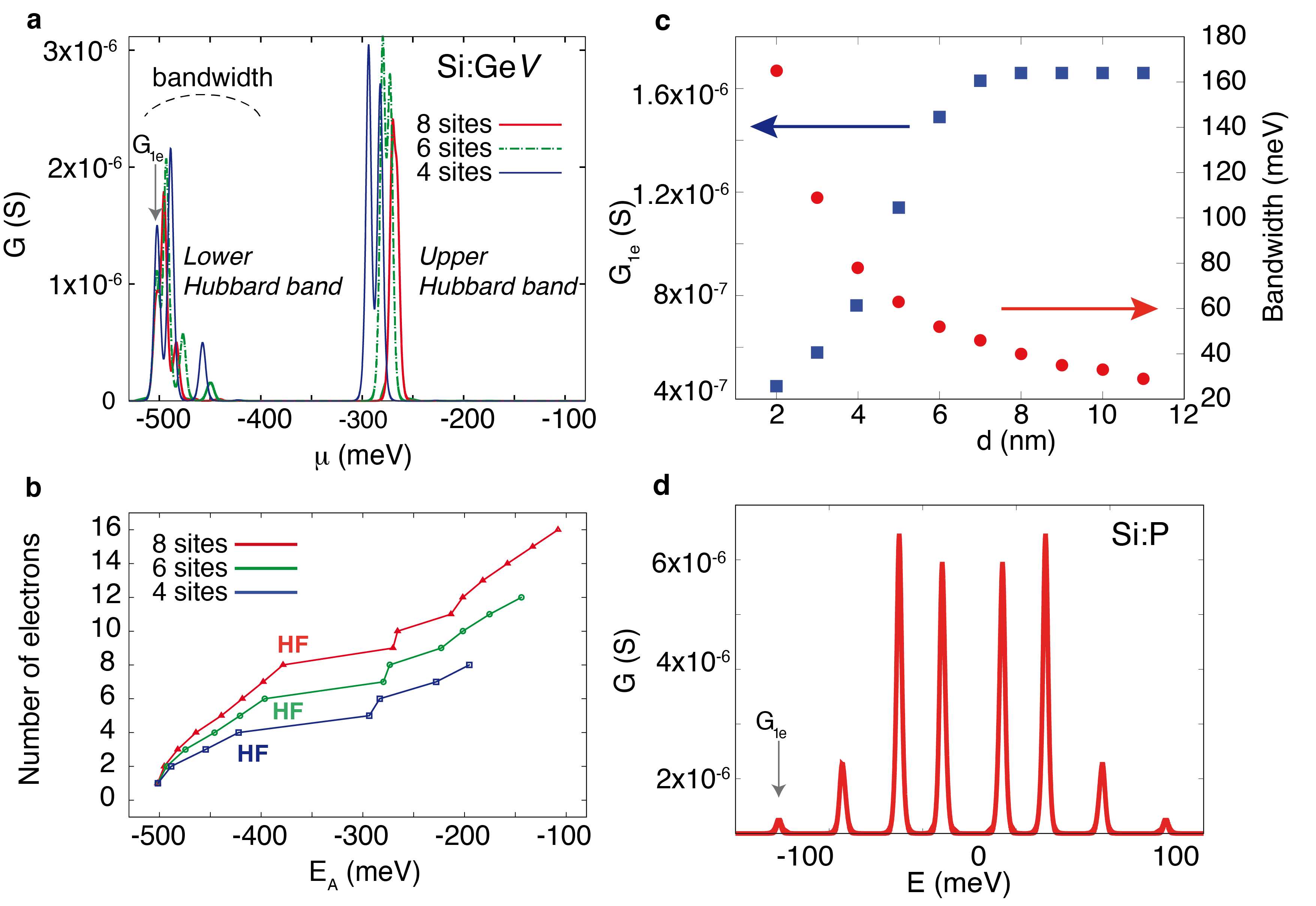}
\caption{\label{panel}
(a)
Computed
conductance at $T$=$20$~K,
as a function of the chemical potential for Ge$V$ arrays consisting of $N_S$=$4$, 6 and 8 sites placed at a regular
spacing of 4~nm. The chemical potential
is referred to the conduction band bottom.
The positions of the peaks match the addition
energies recorded
in the horizontal 
axis of panel (b), obtained at $T$=$0$~K.
The half-filling 
(HF)
configuration for each array is marked.
(c) Computed bandwidth and low-filling conductance 
$G_{\rm 1\,e}$=$G(\mu_{n=1}$)
of an array of 8 Ge$V$ sites as a function of the spacing between the sites.
(d) The conductance of an array of $N_S=4$ P sites in silicon at the same  4~nm spacing and $T$=$20$~K as panel (a).
}
\end{figure}

\begin{figure}
\includegraphics[width=0.8\textwidth]{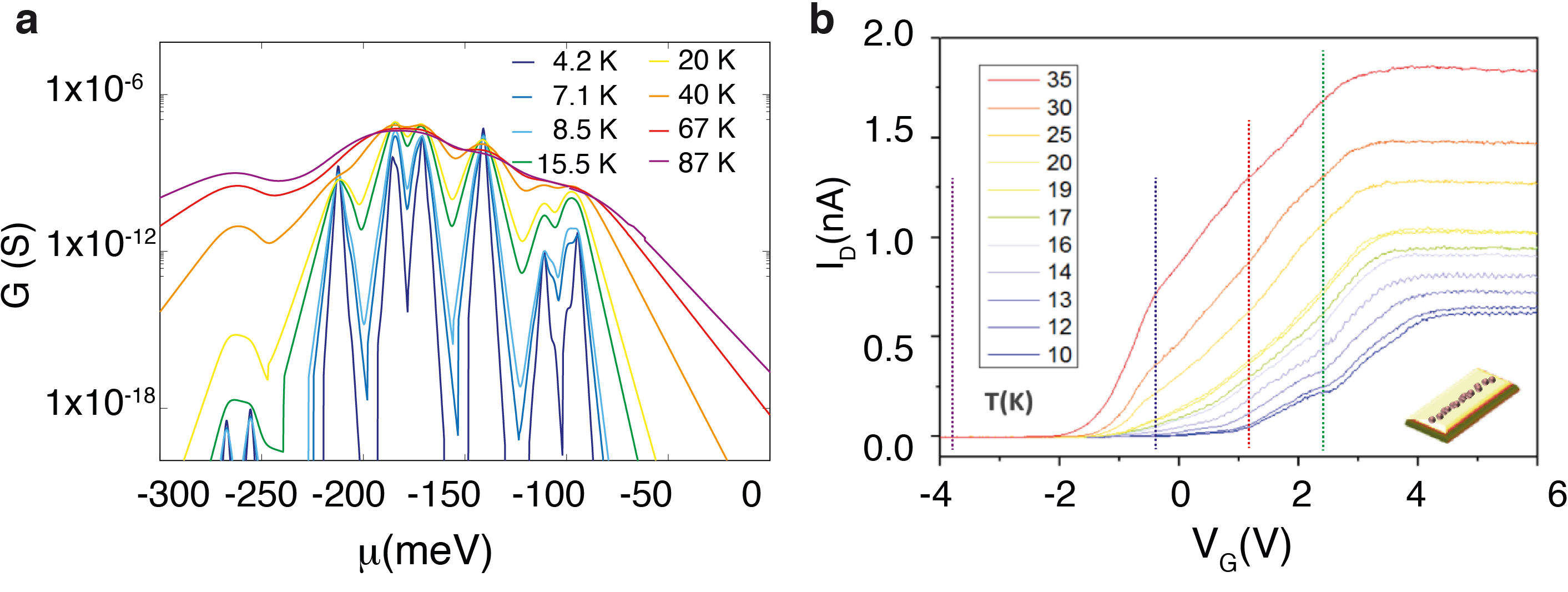}
\caption{\label{short}
(a) Calculated temperature-dependent conductance (upper Hubbard band contribution) of an array of $N_S$=$8$ Ge$V$ complexes at distances 
4$\pm$2~nm.
The chemical potential
is referred to the conduction band bottom.
(b) Trans-characteristics of the short-channel sample with eight pairs of Ge ions implanted by SII.
Blue, red and green vertical dotted lines mark the three sub-bands identified by fitting the data with Fermi Dirac functions (see the text). These three lines and the purple vertical one also mark the gate potential at which the activation energies in Figure~\ref{disord}d and c are extracted.}
\end{figure}

Figure~\ref{panel}a reports the zero-bias conductance
for an ideal array with equally-spaced Ge$V$ sites  placed at a mutual distance of 4~nm,
illustrating the effect of a variable number of sites.

The conductance peaks correspond to the progressive filling of
the $N_S$ sites in the
array with up to $2N_S$ electrons, forming many-body states.

For comparison,
the horizontal axis of Figure~\ref{panel}b reports
the addition energies defined by $E_{ad}(n_e) = E_0(n_e) - E_0(n_e - 1)$, in terms of the ground-state energy $E_0(n_e)$ for $n_e$ electrons.
At $T=0$, the shift in the chemical potential that matches the addition energy leads to the injection of one extra electron
into the channel.

The model exhibits the formation of two Hubbard bands separated by an energy gap which is a result of the on-site electronic correlation.
By increasing the number $N_S$ of sites in the array, the Hubbard gap between these bands
decreases
due to the long-range Coulomb interaction $W_{ij}$.

Indeed the electron-electron repulsion term $W_{ij}$ increases the width of the Hubbard bands that would otherwise be of order $t_{i,j}$, namely very narrow compared to the Hubbard gap.
This bandwidth increases for increasing number of sites $N_S$ (Figure \ref{panel}b), while it decreases by increasing the distances between the defects (Figure~\ref{panel}c),
of course due to the decrease in both $W_{i,j}$ and  $t_{i,j}$.

On the basis of the theoretical description of the Hubbard system we can infer that only the upper Hubbard band is accessible in the experimental measurements, being the lower one very deep in energy.
Although it was not possible to determine the lever arm factor because the transport regime does not involve a Coulomb blockade, the observation of the only upper Hubbard band is compatible with previous reports on the same batch of devices, for which the chemical potential could be tuned by approximately 20 meV per V of applied back-gate voltage \cite{prati2012anderson}. However, the applied voltage may not exceed  
~-10 V, since larger values would lead to dielectric breakdown of the oxide. This implies that the accessible energy range is limited to about 200 meV below the conduction band, not sufficient to reach the lower Hubbard band (500 meV below the conduction band).

In the energy gap of silicon the electric current in the device is driven by electronic states of the electrons injected in the channel.
%
In the ideal array
coherent resonant tunneling \cite{ferraro2015coherent} is carried out by delocalized states in the array, characterized by a sizable
wavefunction amplitude
overlapping the left and right electrodes. 
In contrast,
localized states generated by either Anderson or Mott localization (or both) contribute poorly to the conductance.
These states are expected to contribute to thermally-activated nearest-neighbor
hopping or variable-range hopping \cite{Yu2004}.
On this basis, and 
inspecting
the main contributions to the conductance in Figure~\ref{panel}a, we can infer that
many-body configurations at low-filling of both the lower and upper Hubbard band are
delocalized
across the whole array. 
The large value of the conductance for single-electron filling is a consequence of the degeneracy of the electronic level at the individual sites in the array, due to the neutral character of the Ge$V$ complex.
Consequently, the resonant transport of a single electron
is favored because no localization effects are present.

As the Hubbard band fills up, Coulomb correlations
prevail, causing a drop in conductivity, as can be observed above -450 meV and above -250 meV for the lower and upper Hubbard band, respectively.

At half-filling the system undergoes Mott localization leading to an insulating state with one electron per site.

The quantum transport process in Ge$V$ shows therefore 
significant
differences
compared to the conventional donors.
Being the latter charged impurities, the on-site energies depend on the interaction with the nearby ionic centers, leading to deeper levels at the middle of the array, as shown in the bottom part of Figure~\ref{wavef}b.
As a consequence, in such kind of defects the one-electron transport is disfavored, due to the spatial localization of its ground-state wavefunction at the central sites,
with small overlap with the electrodes.

Indeed, the conductance of a one-dimensional array of P atoms in silicon, obtained with a similar approach \cite{Le} (see Figure~\ref{panel}d),
is maximum near half-filling, where the impurities are neutral, with the result that the electron-electron repulsion ($W_{i,j}$ term) compensates the unequal on-site energies $\epsilon_i$ due to the electron-ion core long-range attraction.

The fingerprint of Mott localization in Ge$V$ is more evident in longer arrays that show a contribution to the conductance only from a small percentage of the associated many-body states (see Figure~\ref{panel}a).
Differently, by keeping constant the number of sites in the 
regularly-spaced
array and increasing their mutual distance, the conductance increases (see Figure~\ref{panel}c)
due to the reduction of the inter-site Coulomb interaction.

In order to 
account for the
main
experimental features,
the ideal assumption of a perfect
regularly-spaced
array has to be relaxed.

For a more realistic model we have included the effect of structural disorder, by introducing a random variability of the inter-site distance, up to $50\%$.
Furthermore, the improved model includes an energy-dependent coupling between the array and the electrodes that takes into account the slower decay of the
wavefunctions of the leads near the top of the
tunneling barrier (see Section~\ref{metods:sec}).

\begin{figure}
\includegraphics[width=1.0\textwidth]{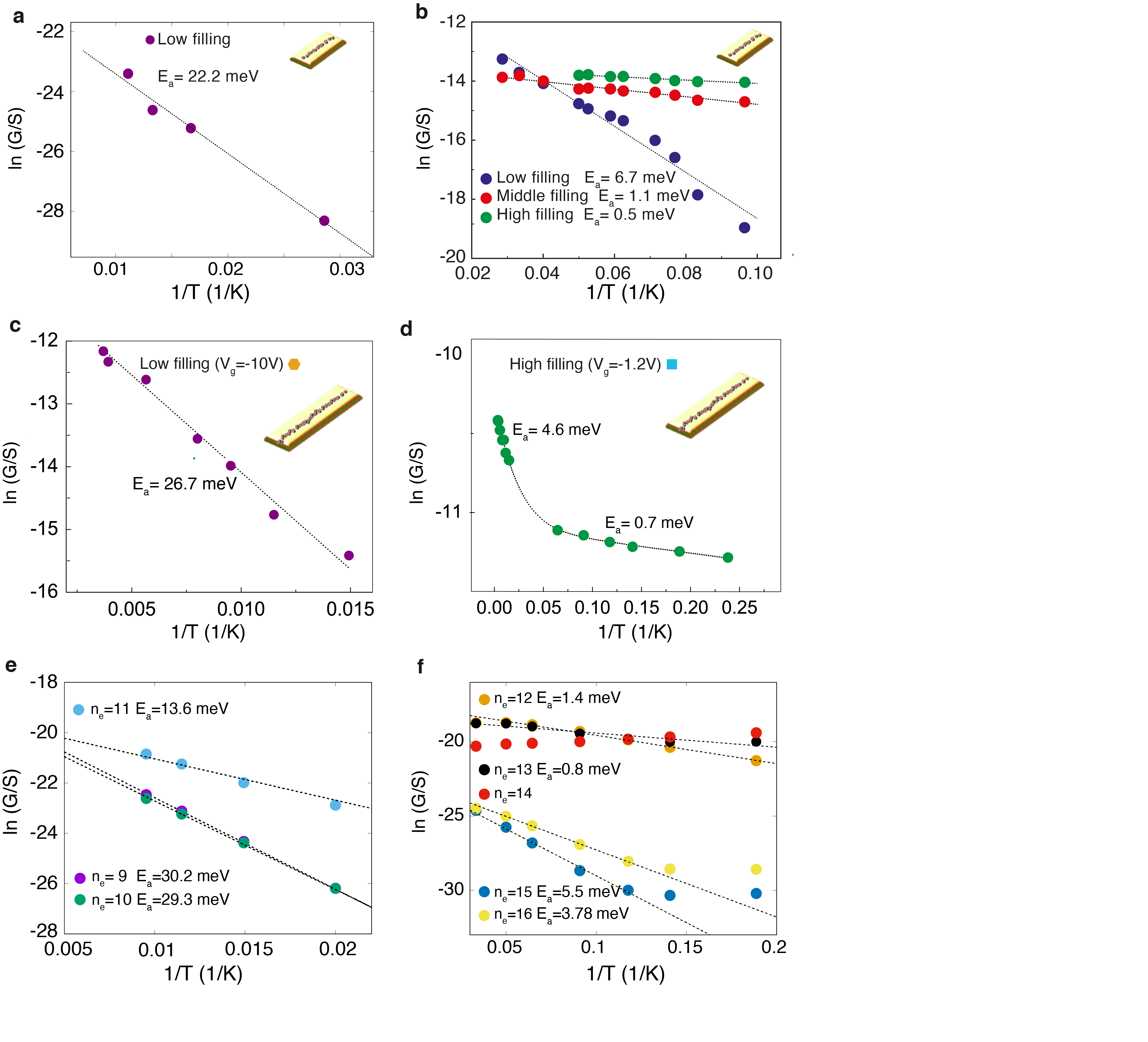}
\caption{\label{disord}
(a) High-temperature dependence of the measured conductance of the short-channel sample at very low filling. The corresponding $V_g$ is marked in Figure~\ref{short}b by a purple vertical line at $\sim$3.8~V.
b) Low-temperature dependence of the measured conductance of the short-channel sample at middle-high filling. The activation energies are extracted at the $V_g$ marked by vertical lines with corresponding color in Figure ~\ref{short}b.
(c) High-temperature dependence of the measured conductance of the long-channel sample in the
low-filling condition (see corresponding $V_g$ in Figure~\ref{Fig-exp1}c).
The activation energies extracted through a fit with the Arrhenius law are reported. 
(d) The same as panel (e) for high-filling (see corresponding $V_g$ Figure~\ref{Fig-exp1}d) and lower-temperature regime.
(e) Arrhenius fits of the
high-temperature dependence of the
theoretical conductivity. The 
individual peaks corresponding to less than a quarter filling of the upper Hubbard band are reported,
labeled by their 
characteristic
values of $n_e$.
(f) The same
as panel a) for the low-temperature regime and high-filling of the upper Hubbard band.
}
\end{figure}

Figure~\ref{short}a reports the computed conductivity for
the upper Hubbard band of
a 1D-array with 8 Ge$V$
sites, spaced randomly in the
2--6~nm range, and temperature from 4.2~K to 87~K.
We focus on the upper Hubbard band because the interpretation of the experimental measurements based on the theoretical results suggest that the lower Hubbard band is too deep in energy to be measured.
Comparison with Figure~\ref{panel}a indicates a sharp reduction in conductivity, which we shall discuss in the next section.

In the disordered array the contribution of states at the middle of the upper Hubbard band ($n_e=12$--14, $\mu\in [-187,-140]$~meV) prevails over the low-filling ones ($n_e=9$--11, $\mu\in [-267,-212]$~meV) as an effect of the energy dependent coupling to the electrodes that hinders the conductivity of deep energy states. The calculated conductance can be compared to the trans-characteristics of the short-channel transistor reported in Figure \ref{short}b for temperatures between 10~K and 35~K and in the Supporting Information for lower and higher temperatures.
For the short-channel sample three sub-bands can be identified and can be related to three different filling regimes, i.e. low, middle and high filling. By fitting the trans-characteristic with three Fermi-Dirac functions we have extracted the $V_G$ corresponding to the onset of the three sub-bands, which is reported in Figure~\ref{short} by blue, red and green lines (see also the Supporting Information). An overall agreement between theory and experiments can be observed,
although the tiny number of defects in the model gives rise to peaks instead of the measured plateau, which is related to the denser spectrum that characterizes channels with many sites.
For what concerns the temperature dependence we observe a general increase of the conductance with the temperature, although in correspondence of some peaks this behavior is not satisfied due to the absence, at these specific energies, of excited states that can be populated by increasing temperature, a condition which is mitigated in long arrays.

By fitting the temperature dependence of the conductance derived from the trans-characteristics in Figures~\ref{Fig-exp1}c,d and \ref{short}b with an Arrhenius law:

$$\ln G(T)=\text{const} - E_a/(k_\text{B}T)$$

we can extract the activation energies $E_a$. We consider two different regimes, i.e.
those
corresponding to very low (less-than-a-quarter) filling (Figure~\ref{disord}a and c) and 
to
middle-high (more-than-a-quarter)
filling (see Figure~\ref{disord}b and d)
We find that at middle-high filling and low temperature (4~K--20~K) the the measured activation energies
of Ge$V$ defects are of the order of 1 meV or less (red and green line of Figure~\ref{disord}b and Figure~\ref{disord}d), i.e. smaller than those of typical donors such as arsenic and phosphorous at low temperature ($\sim 1.2$--$2.5$~meV), as observed in few-dopants arrays.

Also the activation energy at high temperature, i.e. above 20~K (Figure~\ref{disord}d), is smaller than that reported for conventional dopants (4.6~meV versus 7~meV).


In the low-filling regime (see Figure~\ref{disord}a and c) the measured conductance at high temperature is characterized by a very large 
activation energy, of the order of 30 meV.
In particular for the short-channel transistor we found an activation energy of 22.2 meV while for the long-channel one it reaches the value of 26.7 meV.


The theoretical model is able to reproduce the experimental behavior for both the regimes as observed in Figure~\ref{disord}e and f for an array with eight Ge$V$ sites and in the Supporting Information for longer systems including up to twelve sites.
In particular the conductance peaks in the middle of the band ($n_e=12$, $n_e=13$ in Figure~\ref{disord}f) are characterized by activation energies of the order of meV in the temperature range $4$--$20$~K.
Differently, the peaks $n_e=9$ and $n_e=10$ (Figure~\ref{disord}e) display large activation energies at high temperature, reaching 30 meV at the bottom of the upper Hubbard band, in agreement with the experiments.

It is worth noting that despite the theoretical model can be directly compared to the measurements performed on the short-channel sample, which counts approximately the same number of sites in the channel, the peculiar temperature dependence observed for different filling and temperature regimes is reproduced also by increasing the number of sites up to twelve and increasing the distance between the defects (see Supporting Information).

Moreover the statistical analysis,
performed on 10 different random spatial
disorder
realizations
for each theoretical sample with fixed number of sites,
confirms the observed behavior (see Supporting Information).

\section{Discussion}\label{discussion:sec}

The sharp reduction
of the conductivity computed
for the realistic model
compared to the perfect array is due to the
strong
localization of the electrons induced by the structural disorder
enhanced by the large sensitivity of $t_{i,j}$ on the inter-site spacing, a result of the fast decay of the wavefunctions of deep levels.
Disorder is especially effective in 
reducing the intensity of the conductance at low-filling of the upper Hubbard band,
relative to the ideal array.
Consequently, the contribution of extended states is quenched,
with a residual conductance supported by rather localized states, explaining the very low
conductance
in the experiments and the need for a very dense array of implanted complexes to measure a current signal.

Calculations exhibit the largest conductance at the middle of the upper Hubbard band, corresponding to a filling $n_e/N_S \simeq 1.5$.
In this configuration the many-body wavefunction is sufficiently spread over multiple
sites to support
electronic transport.
Indeed this intermediate filling takes the best advantage of the on-site Coulomb repulsion, which tends to favor a leveling of the on-site average occupations, contrasting disorder and long-range Coulomb effects.
As a result, at the middle of the upper Hubbard band
a partial balancing between the localization induced by disorder and the strong Coulomb repulsion that characterizes states at complete filling
supports transport.

The model based on a relatively short array does not grasp all experimental features.
We can expect that simulations of a longer array would provide a larger bandwidth and a denser electronic spectrum, with a fast-growing number of excited states.
These excitations 
are expected to play a significant role in the thermally-activated conductivity.
By increasing $N_S$, the peaks observed in the conductivity at finite $T$ would 
turn into a continuous band and
the conductivity in 
the Hubbard band would appear as a plateau, as in the experiment.

By raising temperature, the peak conductivity increases
due to the thermally-activated
contributions of excited states
delocalized across the entire 1D array.
Despite mismatched array lengths, measurements and calculations agree on indicating that 
at low temperature the conductance 
behaves approximately
as the parallel of two
thermally-activated currents, associated to small activation energies near 1 meV and 5 meV (see Figure~\ref{disord}b,d and f).
This mild temperature dependence of the conductance is likely the result of few delocalized states intersparsed in a dense excited-state energy spectrum of mainly localized strongly-correlated states.

Also the
large activation energy observed at
large negative gate voltage
(Figure~\ref{disord}a and c) is reproduced by the model at 
the beginning of the upper Hubbard band, $n_e/N_S \gtrsim 1$ (Figure~\ref{disord}e).
In this small-filling regime measurements and calculations agree on an activation energy of the order of 30~meV.
It is worth noting that the large activation energy extracted from the calculation for temperatures between 20 and 87~K suggests that
at room temperature
the small-filling contributions
($n_e=9,10$ in Figure~\ref{disord}e)
are indeed expected to reach a value comparable to that of the main peaks, leading to a measurable signal.

This
large activation energy observed in the Ge$V$ 1D array compared
to other dopants can be related to the delocalized nature of the states supporting electronic transport.
At $n_e/N_S \gtrsim 1$, an essentially single electron is pushed across the Ge$V$ array charged by the other $n_e=N_S$ electrons.
The long-range Coulomb repulsion and disorder tend to localize all or almost all its low-energy states.
Delocalized states start to become available beyond an energy of the order of the difference in Coulomb repulsion near the centre and at the ends of the array, i.e.\ for $k_\text{B} T \simeq 30$~meV.
This condition affects less the $n_e/N_S \simeq 1.5$ filling, where the on-site Coulomb repulsion fabricates a few relatively delocalized current-carrying many-body states even at much lower energy.

The model analysis allows us to formulate
two different pictures for the two kind of impurities, namely conventional P dopants and Ge$V$ complexes. 
The former are characterized by a larger intersite hopping,
a smaller on-site repulsion, and neutral energy-aligned sites in the upper Hubbard band.
For these reasons, P or As arrays are less affected by disorder and tend to support larger values of conductance than Ge$V$ arrays. 

On the other hand, in Ge$V$ arrays both the disorder and effective Coulomb-repulsion site-dependent profile are very effective in acting against transport. 
Here localized states prevail by far and conductance increases weakly by raising temperature.
As an exception, at low-filling in excess of half-filling
a few highly-excited
delocalized states can provide a sizable contribution to conduction at room temperature.

\section{Conclusions}

Ge$V$ complexes provide a viable tool to exploit deep levels typical of the vacancy defect in silicon, by means of the precise positioning of single-ion implantation.
By investigating the quantum transport through an array of Ge$V$ complexes in silicon-based transistors, we demonstrate the activation of the Ge$V$ complexes at the low annealing temperature of $550 ^\circ$C and the formation of an impurity band below the conduction-band edge, representing the upper Hubbard band.
The peculiar activation energies of the Ge$V$ complexes induce effects which differ significantly from those of conventional dopants, such as P and As.

By exploiting an
extended Hubbard model based on
{\it ab-initio}-derived parameters,
we explain the thermally-activated contribution to the conductance of weakly-localized states at high filling and low temperature, for which an activation energy of less than 1 meV is observed.
In contrast, at low-filling
we associate fairly delocalized
states to a high activation energy, in the 30~meV region.

We have shown that
the single-ion implantation method enables the engineering of Ge$V$ complexes, a step toward future construction of spatially-controllable single defects in silicon,
with potential exploitation in the control of  quantum information encoded in electron states, with additional interest for
the analog
quantum simulation of strongly-correlated models.

\section{Theoretical model}\label{metods:sec}

We employ an extended Hubbard description of the 1D array of defects in silicon, Eq.~\eqref{model:eq}.
Its parameters are obtained by {\it ab-initio} calculations \cite{achilli-sci-rep}.
The on-site correlation energy is evaluated as the energy separation between the
Ge$V^-$ and Ge$V^{2-}$ charge states, amounting to
$U = 150$~meV.
The other parameters are calculated through the matrix elements, 
whose expression is provided
in Ref.~\cite{Le}.
The involved integrals 
are evaluated
exploiting
{\it ab-initio} wavefunctions computed previously \cite{achilli-sci-rep}, centered on different defects.
The wavefunction of the
in-gap
state
localized around the defect
is available on a real-space grid, whose extension
coincides with
the size of the {\it ab-initio} simulation cell.

In order to account for the
long-distance
decay
of the wavefunction we have 
extended
the size of the simulation cell compared to our previous work \cite{achilli-sci-rep}, 
to a $4\times4\times4$ silicon bulk supercell.
This enlargement of the computational cell has been made possible by adopting a theoretical approach based on a localized basis set and pseudopotential description of the electron-ion interaction, as implemented in the SIESTA package \cite{Sole02}. 
For the exchange-correlation part,
rather than an hybrid functional
we 
adopted
a GGA-1/2 approximation \cite{Ferreira} , which is able to reproduce both the silicon energy gap and the excited states of the defect, in very good agreement with the results we obtained with
the more accurate (but computationally more expensive) hybrid functional \cite{Sko14}.
Being the GGA-1/2 approach not suitable to compute forces, we kept the defect geometry as previously determined in the relaxation done with the hybrid functional \cite{achilli-sci-rep}.
For the calculations of 
the two-centers integrals following Eq.~3-7 of Ref.~\cite{Le}, we extended the wavefunction to larger distances, by extrapolating
its natural decay.

The conductance related to resonant transport is obtained by exploiting the rate equation derived from the Fermi golden rule \cite{Beenakker,Chen,Le}
that yields the tunneling probability of electrons from the left to the right lead, accounting for the transition from a $\Psi_\beta^{n_\sigma-1,n_{\sigma'}}$  to a $\Psi_\alpha^{n_\sigma,n_{\sigma'}}$ many-body state in the array:
\begin{equation}
\begin{aligned}
G=&\frac{2\pi\Gamma e^2}{\hbar k_\text{B}T}\sum_{n_\sigma,n_\sigma'}\sum_{\alpha,\beta} 
\frac{M_{\alpha,\beta,\sigma}^{L,n_\sigma,n_\sigma'} M_{\alpha,\beta,\sigma}^{R,n_\sigma,n_\sigma'}}
{M_{\alpha,\beta,\sigma}^{L,n_\sigma,n_\sigma'}+ M_{\alpha,\beta,\sigma}^{R,n_\sigma,n_\sigma'}}\\
&\times P_i^{n_\sigma,n_\sigma'}[1-f_\text{FD}(E_i^{n_\sigma,n_\sigma'}-E_j^{n_\sigma-1,n_\sigma'}-\mu)]
\,.
\end{aligned}
\end{equation}
Here $E_\alpha^{n_\sigma,n_\sigma'}$ is the energy of the $\alpha$-th many-body state 
with
$n_e=n_\sigma+n_\sigma'$
electrons;
$M_{\alpha,\beta,\sigma}^{L,n_\sigma,n_\sigma'}$ 
($M_{\alpha,\beta,\sigma}^{R,n_\sigma,n_\sigma'}$)
are the transition rates between state $(\alpha,n_\sigma,n_\sigma')$ and $(\beta,n_\sigma\pm1,n_\sigma')$ 
that involve the injection of one electron from (to) the left (right) lead;
$P$ is the the grand canonical ensemble probability that controls the filling of a many-body level at a given temperature;
$f_\text{FD}$ is the Fermi-Dirac distribution;
finally, $n_\sigma$ is the number of the involved up- or down-spin electrons in the array, with $n_\sigma'=n_e-n_\sigma$ those of the other spin kind.
In the evaluations of $G$, 
at very low temperature ($T\leq 3$~K)
in the sums we include only the 
2 lowest-energy many-body states for each 
$(n_\sigma,n_\sigma')$;
at higher temperature 
we include the
10 lowest excited states.

The pre-factor $\Gamma$ includes the coupling to the electrodes, which accounts for the hopping between the leads and the end sites in the array.
For conventional shallow dopants $\Gamma$ is taken as a constant energy of the order of 
a fraction of meV.
In the calculations for the ideal system (Figure \ref{panel}a,d) we fix $\Gamma=0.1$~meV.

The assumption of a constant coupling to the leads should be released for complexes with deep electronic levels, because the wavefunctions of the Fermi-gas electrons in the lead 
decays at a significantly different rate as a function of the energy at which they execute the transition, thus affecting the tunneling overlap with the electrodes.
Accordingly, we modulate the coupling $\Gamma$ with the transmission probability through the potential barrier separating the lead from the channel.
Assuming for simplicity a rectangular potential barrier of height $V_B$ and width $a$, the transmission of an electron at energy $E<V_B$ is given by \cite{Griffiths}:
$$
\vartheta
=\left[1+\frac{V_B^2\sinh^2(ka)}{4E(V_B-E)}\right]^{-1}
$$
where $k=\sqrt{2m (V_B-E)/\hbar)}$.

In
our device
the n-doped silicon forming the leads is smoothly connected to the channel due to the diffusion of P dopants, thus the barrier can be considered extremely
thin.
We 
take $a=a_{Si}~0.34$~nm;
the height of the barrier for the lowest $n_e=1$ state coincides with the excitation energy $0.5$~eV of the ground state of a Ge$V$.
The coupling to the electrodes $\Gamma=\nu\theta$
is set equal to $0.001$~meV 
for the deepest electronic level of the Ge$V$ complex in the lower Hubbard band.
This choice of parameters leads to a computed conductance which is comparable to the experimental one and
the thermally activated behavior is mildly dependent on the choice of $\Gamma$.

\section{Experiment}\label{exp:sec}

Transistors were fabricated on a (100)-oriented n-type (phosphorus $\sim 10^{15}$~cm$^{-3}$) silicon-on-insulator (SOI) wafer by the standard complementary metal-oxide-silicon (CMOS) process (Figure~1).
The source and drain regions are doped with phosphorus to form the n$^+$ diffusion layer.
The channel region (500~nm $\times$ 100~nm in the long-channel sample and 500~nm $\times$ 100~nm in the short-channel one) is exposed to the 
implantation that
can be performed
as the last step
of the transistor fabrication process.
Using the single-ion implantation technique \cite{shinada_2005, ohdomari_2008}, germanium ions were implanted in pairs into the channel region to form a one-dimensional array with 10~nm intervals between
neighboring
sites.
Higher spacing of 40 nm pitch between the implantation sites returned no sub-threshold band.
The implantation of
pairs of ions
increases indeed the probability to have at least one activated complex per site.
The residual randomness of the position of the atoms implanted by SII may cause difficulties for a straightforward implementation of single atom-based  devices. This effect could be mitigated in those application where a small cluster of impurities is sufficient instead of a single atom. Combination of masks with SII can be also considered.
Different annealing temperatures have been applied after the implantation process and we observe that the Ge-related complexes are activated by annealing at 550$^\circ$C for 1~min in N$_2$ gas, as previously reported  \cite{Supr95,schulz_1974}.
 
\begin{figure}
\centering
\includegraphics[width=0.5\textwidth]{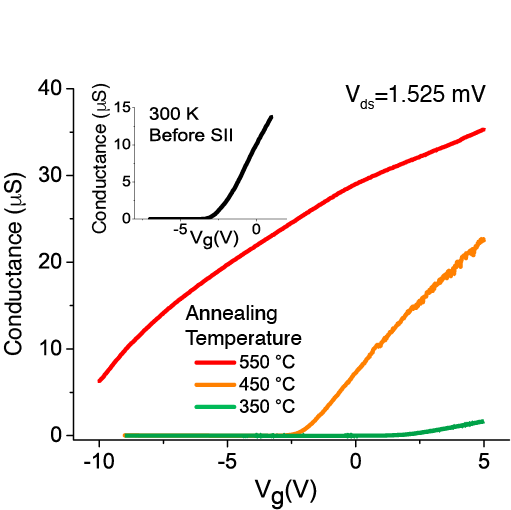}
\caption{\label{Figure5}
    The trans-characteristics of an implanted nanotransistor at three different annealing temperature of $350^\circ$C
    (green),
    $450^\circ$C (orange) and $550^\circ$C (red).
    Transport is restored at $450^\circ$C (to be compared with the inset, which shows a device as before implantation), while the Ge$V$ defects are activated only above $550^\circ$C as from previous reports.
    All conductance measurements have been carried out at room temperature.
}
\end{figure}

In order to demonstrate the activation of the defects at 550$^\circ$C, a set of devices have been prepared to be annealed at different temperature.
Figure~\ref{Figure5} shows the trans-characteristics of the transistor both before the implantation (in the inset) and after annealing
at three different
temperatures.
After the implantation, damages in the crystal significantly reduce the mobility of the electrons.
The annealing at $350^\circ$C is insufficient to restore the lattice.
By the annealing at 450$^\circ$C, the transport of the transistor
conductance
is almost restored, even though
with a  lower mobility
than the
pre-implantation one.
%
As known from the literature  \cite{Supr95,schulz_1974}
such temperature is insufficient for the activation of the Ge$V$ defect, thus here Ge acts by just affecting the mobility.
Finally, by applying an annealing temperature of 550$^\circ$C the Ge$V$ defects are activated, as demonstrated by the formation of an impurity band.    

The Ge$V$ generation in silicon, according to transport data, looks at least of the same order than formation of Ge$V$ in diamond for which the probability to have a single emitter over 200 implanted Ge ions per spot was estimated equal to 53\% \cite{NPJ2018}.

\textbf{Supporting Information} \par Supporting Information is available from the Wiley Online Library or from the author.

\medskip
\textbf{Acknowledgements} \par 
The Authors acknowledge financial support of the NFFA infrastructure under Project ID 517. Computational resources were provided by the INDACO Platform
for
High Performance Computing at the Universit\`a degli Studi di Milano \url{http://www.indaco.unimi.it}, and by CINECA through the 
supercomputing grant project HP10C3S9Z0.
EP, TS and TT acknowledges JSPS, Ministero Affare Esteri, MEXT and CNR Short Mobility program for funding.


\medskip
\bibliographystyle{MSP}
\bibliography{biblio}

%



\begin{figure}
\textbf{Table of Contents}\\
\medskip
  \includegraphics[width=5.5 cm]{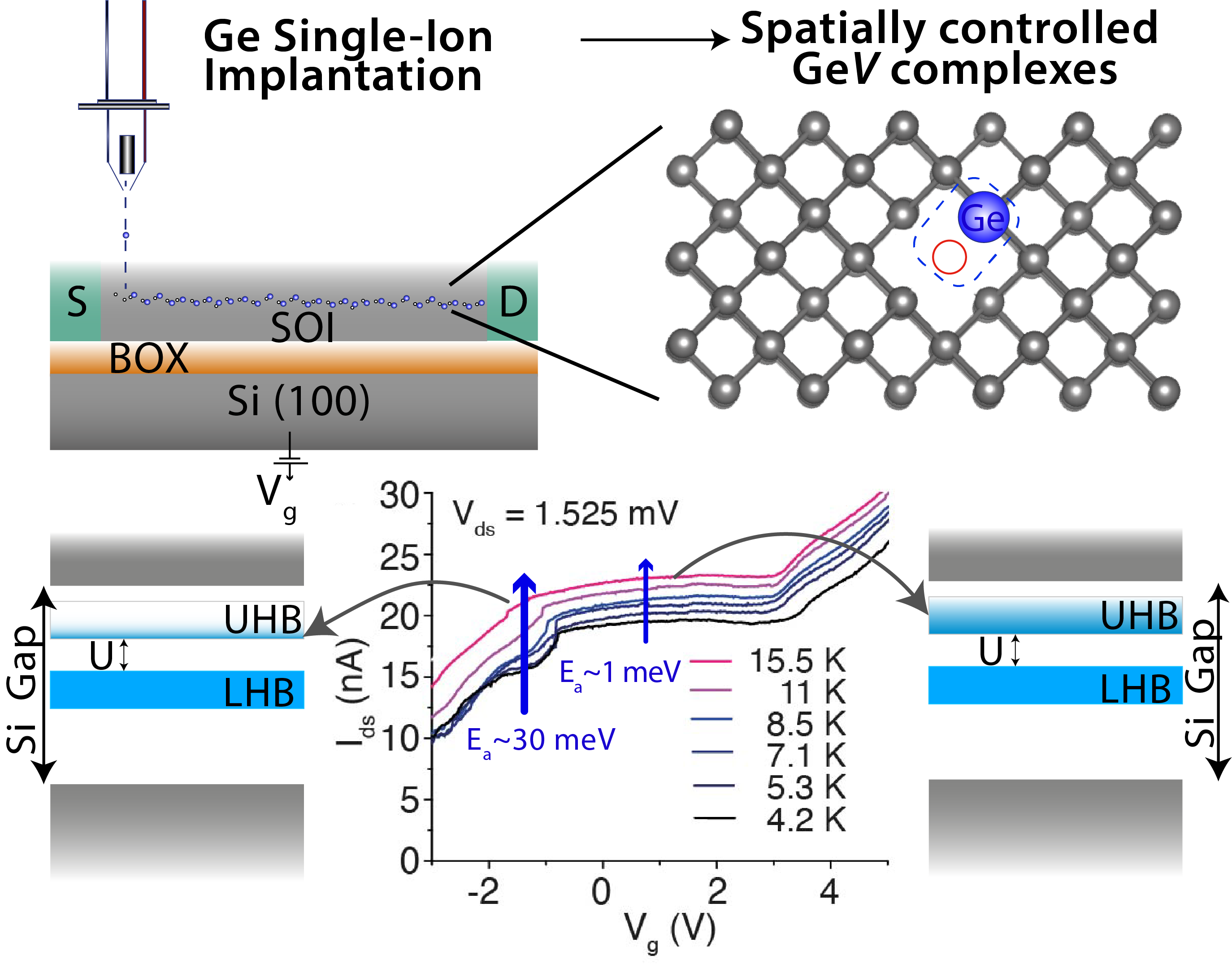}
  \medskip
  \caption*{Spatially controlled functionalized vacancy are created in a silicon channel via high-precision single-ion implantation. Upon annealing they form point defects with deep states in the silicon gap operating up to room temperature. Quantum transport through an array of Ge$V$ complexes displays contributions from weakly localized and delocalized states in the upper Hubbard band, characterized by very different activation energy ($E_A$).}
\end{figure}

\end{document}